\documentclass[12pt,epsf]{article}
\usepackage{graphicx}
\usepackage{float} 
\usepackage{amsmath,latexsym,amssymb,amsfonts}
\usepackage{mathrsfs,array,multirow}
\usepackage{color}
\usepackage{cite}
\usepackage{mathrsfs}
\usepackage{amsmath}
\usepackage{amssymb}
\usepackage{amstext}
\usepackage{color}
\usepackage{subfigure}
\usepackage{epstopdf}
\usepackage{hyperref}     % Internal and external hyper links
\usepackage{cancel}
\usepackage{soul}
\hypersetup{ colorlinks   = true, % Color links instead of ugly boxes
urlcolor     = green, % Color of external hyperlinks
linkcolor    = blue, % Color of internal links
citecolor    = red }	 % Color of citations 
\newcommand{\be}{\begin{equation}}

\newcommand{\ee}{\end{equation}}

\newcommand{\ba}{\begin{array}}

\newcommand{\ea}{\end{array}}

\begin{document}

\begin{titlepage}

%---------------- preprint number ---------------
\hfill\parbox{5cm} { }

\vspace{25mm}

\begin{center}
%------------------------ title ------------------------
{\Large\bf Generalized Holographic Cosmology: low-redshift observational constraint}

%---------------- authors and addresses ----------------
\vskip 1. cm
  {
  Sunly Khimphun$^{a, }$\footnote{e-mail : khimphun.sunly@rupp.edu.kh},
  Bum-Hoon Lee$^{b,c,}$\footnote{e-mail : bhl@sogang.ac.kr}, and
  Gansukh Tumurtushaa$^{c,d,}$\footnote{e-mail : gansuh@ntu.edu.tw}}

{\it $^a\,$ Graduate School of Science, RUPP, Cambodia 12150}\\
{\it $^b\,$ Department of Physics, Sogang University, Seoul, Korea 121-742} \\
{\it $^c\,$ Center for Quantum Spacetime (CQUeST), Sogang University, Seoul, Korea 121-742} \\
%{\it $^c\,$ Center for Theoretical Physics of the Universe, Institute for Basic Science (IBS), Daejeon 34051, Korea}\\
{\it $^d\,$ Leung Center for Cosmology and Particle Astrophysics (LeCosPA), National Taiwan University, Taipei 10617, Taiwan, ROC}

\vskip 0.5cm

%\today
\end{center}

\thispagestyle{empty}

%----------------------- abstract ----------------------

%\centerline{\bf ABSTRACT} \vskip 4mm
\abstract

Four-dimensional cosmological models are studied on a boundary of a five-dimensional Anti-de Sitter ($AdS_5$) black hole with AdS Reissner-Nordstr$\ddot{\text{o}}$m and scalar charged Reissner-Nordstr$\ddot{\text{o}}$m black hole solutions, where we call the former a ``Hairless" black hole and the latter a ``Hairy" black hole.  To obtain the Friedmann-Robertson-Walker (FRW) spacetime metric on the boundary of the $AdS_5$ black hole, we employ Eddington-Finkelstein (EF) coordinates to the bulk geometry. We then derive modified Friedmann equations on a boundary of the $AdS_5$ black hole via AdS/CFT correspondence and discuss its cosmological implications. The late-time acceleration of the universe is investigated in our models. The contributions coming from the bulk side is treated as dark energy source, and we perform MCMC analyses using observational data. Compared to  the $\Lambda$CDM model, our models contain additional free parameters; therefore, to make a fair comparison, we use the Akaike information criterion (AIC) and the Bayesian information criterion (BIC) to analyze our results.  Our numerical analyses show that our models can explain the observational data as reliable as the $\Lambda$CDM model does for the current data. 

\vspace{1cm}

\vspace{2cm}
%PACS numbers:

%\today

\end{titlepage}

%\tableofcontents

\section{Introduction}

There are many dark energy models having been widely studied, which can be categorized as the models of $\Lambda$CDM, quintessence~\cite{Zlatev:1998tr, Steinhardt:1999nw, Zhang:2005rg, Zhang:2005rj}, Chevalliear-Polarski-Linder (CPL) \cite{Chevallier:2000qy, Linder:2002et}, holographic principle \cite{Li:2004rb, Wei:2007ty, Gao:2007ep, Cai:2018ebs}, Dvali-Gabadadze-Porrati (DGP) braneworld \cite{Dvali:2000hr}, and Chaplygin gas model \cite{Kamenshchik:2001cp, Bento:2002ps, Xu:2016grp}. One may refer to Ref.~\cite{Xu:2016grp} for various model comparisons. 
Many of these dark energy models are the theoretical variants of the cosmological constant model, while some are based on totally different theoretical considerations. For example, based on the slowly rolling scalar field, quintessence models produce a negative pressure for the accelerating universe. On the other hand, in the CPL model, the equation-of-state parameter is a function of time. The dark energy models based on quantum gravity theory are often regarded as the holographic models. The models in this category describe the observational data well despite its distinguished theoretical nature to $\Lambda$CDM model. The DGP is also another interesting framework with the realization that higher-dimensional gravity affects the bulk at a large distance from which the dark energy naturally emerges. Another interesting theory is the Chaplygin gas model, which has a connection with the string theory of the braneworld scenario, whose theoretical variant so-called Generalized Chaplygin Gas (GCG) has a well-fitting observational constraint. Another interesting theory is based on Gauge/Gravity theory. It was first introduced in Ref. \cite{Apostolopoulos:2008ru} and further developed by Ref.\cite{Banerjee:2012dw, Camilo:2016kxq} after the boundary metric can be set as dynamical field \cite{Compere:2008us}. In particular, such theory has not yet been fit with with observational constraint. 

In this work, we want to study cosmology from the perspective of holography; in particular, the $AdS/CFT$ correspondence \cite{ Apostolopoulos:2008ru, Banerjee:2012dw, Camilo:2016kxq}. %from which we derive modified Friedmann equations and compare them with the $\Lambda$CDM model. 
The dynamical evolution of the universe in four dimension can be described by the FRW metric, which arises from the boundary of $\text{AdS}_5$ black hole. Starting from the $AdS_5$ black hole geometry, the FRW metric is realized at its four-dimensional boundary via Eddington-Finkelstein (EF) transformation. As a result, four-dimensional gravity with the dynamical FRW metric is foliated since the boundary metric can be stable \cite{Compere:2008us}. This holographic setting is possible mainly based on the idea of mixed-boundary condition studied in Ref.~\cite{Compere:2008us} where such boundary condition allows the boundary metric becoming dynamical.  The black hole as bulk affects the stress-energy tensor due to the AdS/CFT correspondence.  Holographic renormalization is implemented in Ref.~\cite{deHaro:2000vlm}, and for hairy the black hole case, we utilize the counterterm obtained in Ref.~\cite{Papadimitriou:2011qb}. It is worth noting that such a counterterm is obtained in the Fefferman-Graham (FG) coordinate system only. Therefore, the relation between EF and FG coordinate is required.  This mechanism has been studied by Ref.~\cite{ Apostolopoulos:2008ru, Banerjee:2012dw, Camilo:2016kxq}.  In this scenario, we can treat the black hole as a higher dimensional object interacting with an ordinary gravitational theory whose effects play some roles in the cosmological evolution rather than the object where the universe resides.

The cosmological models of our interest are, therefore, extensions to the $\Lambda$CDM model. The reason is that the vacuum energy model in 
four-dimensional gravity theory is foliated at the boundary of one higher-dimensional spacetime. {\color{black} The requirement of such a four-dimensional Einstein-Hilbert action with a cosmological constant is for the purpose of the study of cosmological evolution and due to the consistent form with bare stress-energy tensor on the boundary, which admits the standard interpretation of four-dimensional constant $G_4$ (Newton's constant) and $\Lambda_4$ (cosmological constant) \cite{Banerjee:2012dw}. }

Moreover, we may think of this type of models as a strongly coupled field theory, but as far as an acceleration of the universe is our primary concern, we treat this as a dark energy model.  As an extension, we consider a five-dimensional asymptotically anti-de Sitter(AAdS) black hole with and without a secondary scalar hair and investigate further. We derive the modified Friedmann equation for our models and compare it with $\Lambda$CDM by using observational data, including Supernovae~\cite{Scolnic:2017caz, Alam:2017svz} and $\text{H}_{0}$ measurement data~\cite{ Moresco:2016mzx, Guo:2015gpa}. %After obtaining the modified Friedmann equations at the four-dimensional boundary of five-dimensional bulk spacetime, we adopt the numerical techniques developed in Ref.~\cite{Conley:2011ku, Basilakos:2016nyg,Goliath:2001af, Lazkoz:2007cc, Sanchez:2009ka, Nesseris:2010pc,DeFelice:2012vd} and investigate the dark energy models in light of  the observational data.

%The dynamical evolution of the universe can be described by the FRW metric. Starting from five-dimensional AdS black hole geometry, the FRW metric is realized at its four-dimensional boundary via Eddington-Finkelstein (EF) transformation. As a result, four-dimensional gravity with the dynamical FRW metric is foliated since the boundary metric can be stably dynamical \cite{Compere:2008us}. A black hole as bulk will affect the stress-energy tensor due to the AdS/CFT correspondence. In this scenario, we can treat the black hole as a higher dimensional object interacting with an ordinary gravitational theory whose effect play some roles in the cosmological evolution rather than the object where the universe resides. Holographic renormalization is implemented in Ref.~\cite{deHaro:2000vlm}, and for hairy the black hole case, we utilize the counterterm obtained in Ref.~\cite{Papadimitriou:2011qb}. It is worth noting that such a counterterm is obtained in the Fefferman-Graham (FG) coordinate system only. As a result, the relation between EF and FG coordinate is required.  After obtaining the modified Friedmann equations at the four-dimensional boundary of five-dimensional bulk spacetime, we adopt the numerical techniques developed in Ref.~\cite{Conley:2011ku, Basilakos:2016nyg,Goliath:2001af, Lazkoz:2007cc, Sanchez:2009ka, Nesseris:2010pc,DeFelice:2012vd} and investigate the dark energy models in light of  the observational data.

This paper is organized as follows. In section \ref{AdS5-FRW}, we review a procedure of obtaining the FRW metric at a boundary of the $\text{AdS}_5$ black hole. In section \ref{FriedEq}, we derive modified Friedmann equations by employing the mixed-boundary condition and AdS/CFT correspondence, whose bulk solutions are the charged dilatonic $\text{AdS}_5$ black hole \cite{Gubser:2009qt}. In section \ref{constraint}, we present our numerical fitting results of the MCMC analyses, for which we adopt the numerical techniques developed in Ref.~\cite{Conley:2011ku, Basilakos:2016nyg,Goliath:2001af, Lazkoz:2007cc, Sanchez:2009ka, Nesseris:2010pc,DeFelice:2012vd}. We use the observational data, including Supernova (SnIa) and Hubble expansion rate data~\cite{Scolnic:2017caz, Moresco:2016mzx}, to provide observational bounds on model parameters associated with the late-time dynamics of the universe. %To fit our models to observational data, we use joint analyses of $\chi^2$ statistic, and then will be followed by the comparisons of our model in both cases of hairless and hairy black hole solutions with $\Lambda$CDM model using Akaike information criterion (AIC) \cite{Akaike1974}, and Bayesian information criterion (BIC) \cite{Schwarz:1978tpv}. 
Finally, section~\ref{conclusion} is devoted to summary and conclusions of the present study.

%------------------------------------------------------------------------------------------------------------------------------------------------------------------------------------------------------------------------------
\section{Five-dimensional AdS black hole and FRW boundary}\label{AdS5-FRW}
%------------------------------------------------------------------------------------------------------------------------------------------------------------------------------------------------------------------------------

The idea of realizing the FRW universe at the boundary of the AdS$_5$ black hole was first introduced in \cite{Apostolopoulos:2008ru}, where the Schwarzschild solution was considered, and subsequent works were done in \cite{Banerjee:2012dw, Camilo:2016kxq}. If one can obtain a preferred boundary geometry from the AdS black hole solution, then the concept of having the mixed-boundary condition is essentially required in order to generate a dynamical FRW metric. It was shown that such a boundary condition is dynamically stable \cite{Compere:2008us}. Since the new effective method used in \cite{Camilo:2016kxq} allows one to consider a class of complicated AdS black hole, in this section, we will recap such the method and consider a charged AdS dilatonic black hole solution (hairy black hole). We begin with the general metric 
\begin{equation}
ds^2=-f(r)dt^2+g(r)dr^2+\Sigma(r)^2d\Omega_3^2.
\end{equation}
This metric describes AAdS$_5$ where $f(r)\sim r^2/L^2$, $g(r)\sim L^2/r^2$, and $\Sigma(r)\sim r/L$ for large $r$ at the asymptotic region where $L$ is the AdS radius. Introducing a new coordinates $v$ such that $dt=\pm dv/\sqrt{f(r)g(r)}\mp dr\sqrt{g(r)/f(r)}$, we obtain a metric in the EF coordinates
\begin{equation}\label{EF}
ds^2=2dvdr-f(r)dv^2+\Sigma(r)^2d\Omega_3^2,
\end{equation}
which has the four-dimensional conformal boundary. We adopt a new time and radial coordinates $V$ and $R$, respectively, such that $dv= dV/a(V)$ and $R= r/a(V)$, where $a(V)$ will be the scale factor. Then the metric (\ref{EF}) becomes
\begin{equation}\label{EFmetric}
ds^2=2dVdR-\left[\frac{f(Ra)}{a^2}-2R\frac{\dot{a}}{a}\right]dV^2+\Sigma(Ra)^2d\Omega_3^2\,,
\end{equation}
where the dot represents the derivative with respect to $V$.  In this holographic approach to cosmology, we need to put the boundary hypersurface at a finite distance $R$ with an appropriate counterterm. As can be seen from (\ref{EFmetric}), when large $R$ is fixed, the boundary metric reduces to FRW metric as desired. The EF coordinates associating with new time and radial coordinates is not well-understood in holographic renormalization context. For this reason, we need to find the relation between the EF and FG coordinates which is given by
\begin{equation}\label{FG}
ds^2=\frac{L^2}{z^2}\left[dz^2+g_{\mu\nu}dx^{\mu}dx^{\nu}\right]\,,
\end{equation}
where
\begin{equation}\label{metric}
g_{\mu\nu}(z,x)=g_{\mu\nu}^{(0)}(x)+z^2g_{\mu\nu}^{(2)}(x)+z^4\left(g_{\mu\nu}^{(4)}(x)+h^{(4)}_{\mu\nu}(x)\log z\right)+\cdots\,,
\end{equation}
is defined as an appropriate form of ansatz for Fefferman-Graham asymptotic expansion \cite{deHaro:2000vlm}. Comparing the metric (\ref{EF}) with (\ref{FG}), we obtain the following two relations
\begin{equation}\label{PD}
\begin{split}
&2\partial_zR\partial_zV-\alpha(\partial_z V)^2=\frac{L^2}{z^2},\\
&\partial_zV\partial_{\tau}R+\partial_{\tau}V\partial_{z}R-\alpha\partial_{z}V\partial_{\tau}V=0\,,
\end{split}
\end{equation}
where $\alpha= f(Ra)/a^2 - 2R \dot{a}/a$. The boundary metric can be obtained in the same way as
\begin{equation}\label{metric}
g_{\tau\tau}=-\frac{(\partial_{\tau}V)^2}{(\partial_{z}V)^2}, ~~ g_{ij}dx^idx^j=\frac{z^2}{L^2}\Sigma^2(Ra)d\Omega_3^2.
\end{equation}
The power series expansion of $V(\tau,z)$ and $R(\tau,z)$ will be obtained using (\ref{PD}) and the metric $g_{\mu\nu}$written in terms of $z$ and $\tau$ will be determined by (\ref{metric}).

The equation governing the cosmological evolution can be derived by using the Friedmann equation. In our study, we will obtain the modified Friedmann equations due to the contribution from higher dimension via AdS/CFT correspondence. In other  words, the modified terms come from the regulated stress-energy tensor of the dual conformal field theory residing on the four-dimensional boundary hypersurface. Adopting the mixed-boundary condition, we can write our action as the following
{\footnotesize 
\begin{equation}\label{eq:generalaction}
S=\int_{\mathcal{M}}d^5x\sqrt{- \text{det} g_5}\mathcal{L}^{\text{gravity}}_{5\text{D}}+\frac{1}{16\pi G_4}\int_{\partial \mathcal{M}}d^4x\sqrt{-\text{det} g^{(0)}}(R-2\Lambda_4)+\int_{\partial \mathcal{M}}d^4x\sqrt{-\text{det} g^{(0)}}\mathcal{L}^{\text{matter}}_{4\text{D}}\,,
\end{equation}}
where $g_5$ is a five-dimensional metric. We define $\mathcal{L}^{\text{gravity}}_{5\text{D}}$ as the Lagrangian representing the five dimensional Einstein gravity with negative cosmological constant. Notice that $g^{(0)}$ is the leading order in metric of a four dimensional boundary hypersurface corresponding to FG coordinate introduced in (\ref{metric}). Mixed-boundary condition implies that the total stress-energy tensor, which include $T^{\text{CFT}}_{\mu\nu}$, $T^{\text{4D}}_{\mu\nu}$, and $T^{\text{matter}}_{\mu\nu}$ is zero so that the variational principle still holds \cite{Compere:2008us, Banerjee:2012dw}. Thus, the five dimensional dual field theory will contributes to the stress-energy tensor and modified the equation of motion. Finally, we define $\mathcal{L}^{\text{matter}}_{4\text{D}}$ as the Lagrangian from ordinary matter. The variation of this action gives
\begin{equation}\label{EFE}
R_{\mu\nu}-\frac{1}{2}g_{\mu\nu}^{(0)}R+\Lambda_4g_{\mu\nu}^{(0)}=8\pi G_4\left(\left<T^{\text{CFT}}_{\mu\nu}\right>+T^{\text{matter}}_{\mu\nu}\right).
\end{equation}
The Ricci tensor and scalar are calculated from the zeroth order boundary metric $g_{\mu\nu}^{(0)}$. The stress-energy tensor $\left<T^{\text{CFT}}_{\mu\nu}\right>$ and $T^{\text{matter}}_{\mu\nu}$ are obtained from the AdS$_5$/CFT$_4$ correspondence and from four-dimensional gravity theory with cosmological constant, respectively. The stress-energy tensor from the dual field has conformal anomaly since the boundary has an even dimension. This conformal anomaly will be remedied by the holographic renormalization.

\section{Modified Friedmann equations in a $\text{AdS}_5$ black hole}\label{FriedEq}
In this section, we begin by briefly introducing the five-dimensional scalar charged AdS black hole solution \cite{Gubser:2009qt, Kim:2016dik, Jeong:2019csr}  with the following Lagrangian in (\ref{eq:generalaction}), 
\begin{align}\label{eq:Action}
%&S=\frac{1}{2\kappa^2}\int d^5x\sqrt{-g}\left[R-W(\phi)F^2-\frac{1}{2}(\partial\phi)^2-V(\phi)\right]\,,\nonumber\\
&\mathcal{L}_{\text{5D}}^{\text{gravity}} = R-W(\phi)F^2-\frac{1}{2}(\partial\phi)^2-V(\phi)\,,
\end{align}
with a potential $V(\phi)$ and a coupling $W(\phi)$ of the form
\begin{equation}
\begin{split}
V(\phi)=-\frac{1}{L^2}(8e^{{\phi}/\sqrt{6}}+4e^{-2{\phi}/\sqrt{6}}),~~ W(\phi)=\frac{1}{4}e^{2\phi/\sqrt{6}}\,.
\end{split}
\end{equation}
Scalar field which non-minimally coupled to guage field is considered here because minimal coupling scalar field will result in the trivial solution. As a result, it is out of our interest. A scalar charged Reissner-Nordstr$\ddot{\text{o}}$m black hole solution of the given action is 
\begin{equation}
ds^2=e^{2C}(-hdt^2+d\vec{x}^2)+\frac{e^{2D}}{h}dr^2
\end{equation}
where
\begin{equation}\label{blh1}
\begin{split}
&C=\log\left(\frac{r}{L}\right)+\frac{1}{3}\log\left(1+\frac{Q^2}{r^2}\right),~~
D=-\log\left(\frac{r}{L}\right)-\frac{2}{3}\log\left(1+\frac{Q^2}{r^2}\right),\\
&h=1-\frac{M L^2}{(Q^2+r^2)^2},~~
\phi=\frac{2}{\sqrt{6}}\log\left(1+\frac{Q^2}{r^2}\right),~~
A=\left(-\frac{Q\sqrt{2M}}{Q^2+r^2}+\frac{Q\sqrt{2M}}{Q^2+r_h^2}\right)dt.
\end{split}
\end{equation}
Here $Q$ is the charge and $M$ is the mass of the black hole. The horizon $r_h$ is defined such that $h(r_h)=0$. By introducing a new coordinate defined as follow
\begin{align}\label{dt}
dt %&=-e^{-(C+D)}dv+\frac{e^{(D-C)}}{h}dr\nonumber\\
&=\left[\frac{\dot{a}L^2}{a^2R\left(1-\frac{M L^2}{(Q^2+a^2R^2)^2}\right)\left(1+\frac{Q^2}{a^2R^2}\right)}-\frac{1}{a}\left(1+\frac{Q^2}{a^2R^2}\right)^{\frac{1}{3}}\right]dV\nonumber\\
&\qquad \qquad \qquad \quad \,\,\, +\left[\frac{aL^2}{a^2R^2\left(1-\frac{M L^2}{(Q^2+a^2R^2)^2}\right)\left(1+\frac{Q^2}{a^2R^2}\right)}\right]dR\,.
\end{align}
Also, one can transform the metric into the EF coordinates of the form
\begin{equation}
ds^2=2dvdr-he^{-2D}dv^2+e^{2C}d\vec{x}^2.
\end{equation}
Comparing with the metric expression given in (\ref{EF}), we have
\begin{equation}
f(r)=he^{-2D} \indent \text{and} \indent \Sigma(r)=e^C.
\end{equation}
The next step is to transform this EF coordinates to the FG coordinates for the sake of holographic renormalization. In order to do that, we first consider a power series expansion of coordinates $R$ and $V$ near $z=0$ regime which reads 
\begin{equation}
%\begin{split}
V(\tau,z)=\sum_{n=0}V_{(n)}z^n,  ~~  R(\tau,z)=\sum_{n=0}R_{(n)}z^{n-1}.
%\end{split}
\end{equation}
The followings are coefficients (up to fifth order) obtained by calculating ($\ref{PD}$) with the given black hole geometry order by order,
\begin{align}\label{V}
V_{(0)}&=\tau, \quad V_{(1)}=-1, \quad V_{(2)}=0, \quad V_{(3)}=\frac{-6 a \ddot{a}+3 \dot{a}^2+4 Q^2}{36 a^2}, \nonumber \\
V_{(4)}&=\frac{3 a^2 a^{(3)}+3 \dot{a}^3+\dot{a} \left(4 Q^2-6 a \ddot{a}\right)}{72 a^3}, \\ 
V_{(5)}&=-\frac{6 a^3 a^{(4)}+24 a^2 \ddot{a}^2-40 Q^2 a \ddot{a}-9 \dot{a}^4-18 a^2  \dot{a} a^{(3)}+6 a \dot{a}^2 \ddot{a}+54 M +4 Q^4}{720 a^4}\,.\nonumber
\end{align}
$V_{0}=\tau$ is chosen because $V$ becomes a time $\tau$ at the boundary.
Similarly, the expansion for $R$ starts from $1/z$ and $R_{(0)}=L^2$ since $z\sim L^2/R$ near the boundary. 
\begin{align}
R_{(0)}&=1, \quad R_{(1)}=\frac{ \dot{a}}{a}, R_{(2)}=\frac{-6 a \ddot{a}+9 \dot{a}^2-4 Q^2}{12 a^2}, \quad R_{(3)}=\frac{3 a^2 a^{(3)}+12 \dot{a}^3-\dot{a} \left(15 a \ddot{a}+8 Q^2\right)}{18 a^3},\label{R}  \\ 
R_{(4)}&=\frac{-3 a^3 a^{(4)}+6 a^2 \ddot{a}^2+20 Q^2 a \ddot{a}+39 \dot{a}^4+21 a^2  \dot{a} a^{(3)}-\dot{a}^2 \left(63 a \ddot{a}+44 Q^2\right)+9 M -2 Q^4}{72 a^4}\,,\nonumber
\end{align}
where and hereafter we set $L=1$ for simplicity.
The metrics in the FG coordinates is computed (up to fourth order) applying with a profile of $V$ and $R$ in (\ref{R}) and (\ref{V}) to (\ref{metric})
\begin{align}
&g^{(0)}_{\tau\tau}=-1, \quad g^{(2)}_{\tau\tau}=\frac{6 a \ddot{a}-3 \dot{a}^2-4 Q^2}{6 a^2},\nonumber\\
&g^{(4)}_{\tau\tau}=\frac{48 Q^2 a \ddot{a}-36 a^2 \ddot{a}^2-9 \dot{a}^4+\dot{a}^2 \left(36 a \ddot{a}-24 Q^2\right)-40 Q^4+108 M }{144 a^4},\\
&g^{(0)}_{ij}dx^idx^j=a^2d\Omega_3^2, \quad g^{(2)}_{ij}dx^idx^j=-\frac{1}{2} \dot{a}^2d\Omega_3^2, \quad g^{(4)}_{ij}dx^idx^j=\frac{24 Q^2 \dot{a}^2+9 \dot{a}^4-8 Q^4+36 M }{144 a^2}d\Omega_3^2\,.\nonumber
\end{align}
The zeroth order metric is the FRW metric as intended.

The expansion of  scalar field in terms of ($\tau$, $z$) is
\begin{equation}
\phi(\tau,z)=\sum_{n=0}\phi_{(n)}(\tau)z^{n}.
\end{equation}
Coefficients are obtained using the scalar field given in (\ref{blh1}) and listed in the following (up to sixth order).
\begin{align}
\phi_{(0)}&=\phi_{(1)}=\phi_{(3)}=\phi_{(5)}=0,\nonumber\\
\phi_{(2)}&=\frac{\sqrt{6} Q^2}{3a^2}, \quad \phi_{(4)}=\frac{Q^2 \left( 3  \dot{a}^2+Q^2\right)}{3 \sqrt{6} a^4}, \quad  \phi_{(6)}=\frac{Q^2 \left(27 \dot{a}^4-36 M +8 Q^4\right)}{72 \sqrt{6} a^6}\,.
\end{align}
Note that the zeroth order term vanishes. This is an obvious result since any field having a dual operator should vanish at the boundary.

The one form $dt$ in the gauge field $A$ in (\ref{blh1}) can be converted to the EF coordinates using (\ref{dt}). Then the gauge field becomes
\begin{equation}
A=A_{V}dV+A_{R}dR,
\end{equation}
where
\begin{align}
A_V&=\left(-\frac{Q\sqrt{2M}}{Q^2+a^2R^2}+\frac{Q\sqrt{2M}}{Q^2+r_h^2}\right)\left[\frac{\dot{a}L^2}{a^2R\left(1-\frac{M L^2}{(Q^2+a^2R^2)^2}\right)\left(1+\frac{Q^2}{a^2R^2}\right)}-\frac{1}{a}\left(1+\frac{Q^2}{a^2R^2}\right)^{\frac{1}{3}}\right],\\
A_R&=\left(-\frac{Q\sqrt{2M}}{Q^2+a^2R^2}+\frac{Q\sqrt{2M}}{Q^2+r_h^2}\right)\left[\frac{L^2}{a R^2\left(1-\frac{M L^2}{(Q^2+a^2R^2)^2}\right)\left(1+\frac{Q^2}{a^2R^2}\right)}\right]\,.
\end{align}
The gauge field in the FG coordinates can be obtained easily from the expression in the EF coordinates.
\begin{align}
A&=A_{V}\left(\frac{\partial V}{\partial \tau}d\tau+\frac{\partial V}{\partial z}dz \right)+A_{R}\left(\frac{\partial R}{\partial \tau}d\tau+\frac{\partial R}{\partial z}dz\right) \equiv A_{\tau}d\tau+A_{z}dz \,.
\end{align}
The power series expansions of $V$ and $R$ with respect to $z$ gives the expansion solution for $A_{\tau}$ and $A_{z}$ of the form
\begin{equation}
A_{\tau}(\tau,z)=\sum_{n=0}A_{\tau}^{(n)}(\tau)z^n, ~~
A_{z}(\tau,z)=\sum_{n=0}A_{z}^{(n)}(\tau)z^n.
\end{equation}
Coefficients are given as (up to fourth order)
\begin{align}
A_{\tau}^{(0)}&=-\frac{Q \sqrt{2M} }{a \left(Q^2+r_h^2\right)}, \quad A_{\tau}^{(1)}=0, \quad A_{\tau}^{(2)}=\frac{Q \sqrt{2M } \left(3 a \ddot{a}-6 \dot{a}^2+6 r_h^2+4 Q^2\right)}{6 a^3 \left(Q^2+r_h^2\right)}, \nonumber \\
A_{\tau}^{(3)}&=0\,, \quad A_{\tau}^{(4)}=\frac{Q \sqrt{2M } \left(-36 r_h^2 a  \ddot{a}-36 Q^2 a \ddot{a}+108 r_h^2 \dot{a}^2-36 \dot{a}^4+120 Q^2 \dot{a}^2+27 a \dot{a}^2 \ddot{a}-8 Q^4\right)}{72 a^5 \left(Q^2+r_h^2\right)},\nonumber \\
A_{z}^{(0)}&=A_{z}^{(2)}=A_{z}^{(4)}=0, \quad A_{z}^{(1)}=\frac{ Q \sqrt{2M} \dot{a}}{a^2 \left(Q^2+r_h^2\right)}, \quad A_{z}^{(3)}=-\frac{ Q \sqrt{2M } \left(6 r_h^2 \dot{a}+8 Q^2 \dot{a}-3 \dot{a}^3\right)}{6 a^4 \left(Q^2+r_h^2\right)},
\end{align}

The holographic renormalization of the Einstein-Dilaton action is needed. In \cite{Papadimitriou:2011qb}, the holographic renormalization of the Einstein-Dilaton-Axion action is considered. The desired renormalized tensor can be obtained by turning off the axion field and including gauge field effect.
\begin{align}\label{st1}
\langle T_{\mu\nu}^{\text{CFT}}\rangle=&\frac{1}{\kappa^2}\left[2\left(g_{(4)\mu\nu}-\text{Tr}g_{(4)}g_{(0)\mu\nu}\right)+(h_{(4)\mu\nu}-\frac{1}{2}\text{Tr}h_{(4)}g_{(0)\mu\nu})-(\text{Tr}g_{(2)}g_{(2)\mu\nu}-\frac{1}{2}\text{Tr} g_{(2)}^2 g_{(0)\mu\nu}) \right]\nonumber\\
&+\frac{1}{2\kappa^2}\left[ D_{(0)\sigma}D_{(0)(\mu}g_{(2)\nu)}^{\;\;\;\;\sigma}-\frac{1}{2}\Box_{(0)}g_{(2)\mu\nu}-\frac{1}{2}D_{(0)\mu}D_{(0)\nu}\text{Tr}g_{(2)}-\partial_{(\mu}\phi_{(0)}\partial_{\nu)}\phi_{(2)}  \right. \nonumber\\
&-\frac{1}{2}g_{(2)\mu\nu}\left(R[g_{(0)}]-\frac{1}{2}\partial_{\sigma}\phi_{(0)}\partial^{\sigma}\phi_{(0)} \right)+\frac{1}{2}g_{(0)\mu\nu}g_{(2)}^{\;\;\sigma\rho}\left(R[g_{(0)}]_{\sigma\rho}-\frac{1}{2}\partial_{\sigma}\phi_{(0)}\partial_{\rho}\phi_{(0)}  \right)\nonumber\\
&\left.-\frac{1}{2}g_{(0)\mu\nu}\left(D_{(0)\sigma}D_{(0)\rho}g_{(2)}^{\;\;\sigma\rho}-\Box_{(0)}\text{Tr}g_{(2)}-\partial_{\sigma}\phi_{(0)}\partial^{\sigma}\phi_{(2)}\right)\right]+\frac{1}{48}F^{(0)}_{\sigma\rho}F_{(0)}^{\sigma\rho}g^{(0)}_{\mu\nu}.
\end{align}
The last term in (\ref{st1}) is responsible for the gauge field. This term is the same with the minimal coupling case as the nonminimal coupling term converges to one at the boundary. The explicit formula for the stress-energy tensor is
\begin{align}
\langle T^{\text{CFT}}_{\tau\tau} \rangle&= \frac{24 L^4 Q^2 \dot{a}^2+9 L^8 \dot{a}^4+8 Q^4+36 L^2 M }{192 \pi  \text{G}_5 L^8 a^4}, \nonumber\\
\langle T^{\text{CFT}}_{ij} \rangle &=\frac{-48 L^4 Q^2 a \ddot{a}+9 L^8 \dot{a}^4+12 L^4 \dot{a}^2 \left(2 Q^2-3 a \ddot{a}\right)+8 Q^4+36 L^2 M }{576 \pi  \text{G}_5 L^8 a^2}\delta_{ij}\,.
\end{align}
The Ward identity for this case is
\begin{equation}
\nabla^{\mu} \left<T_{\mu\nu}^{\text{CFT}}\right>=\left<\mathcal{O}\right>\partial_{\nu} \phi_{(0)}+F_{\mu\nu}^{(0)}\left<J^{\mu}\right>=0,
\end{equation}
where $\mathcal{O}$ is the operator correspond to the dual field $\phi$ and $J$ is the current from the dual gauge field $F$. This relation confirms $\nabla^{\mu} T_{\mu\nu}^{\text{matter}}=0$ combining with (\ref{EFE}).
The energy density and the pressure from these relations read
\begin{equation}
\begin{split}
\langle\rho^{\text{CFT}}\rangle &= \frac{24 L^4 Q^2 \dot{a}^2+9 L^8 \dot{a}^4+8 Q^4+36 L^2 M }{192 \pi  \text{G}_5 L^8 a^4},\\
\langle p^{\text{CFT}} \rangle   &= \frac{-48 L^4 Q^2 a \ddot{a}+9 L^8 \dot{a}^4+12 L^4 \dot{a}^2 \left(2 Q^2-3 a \ddot{a}\right)+8 Q^4+36 L^2 M }{192 \pi  \text{G}_5 L^8 a^2}.
\end{split}
\end{equation}
The $\tau\tau$-component of the Einstein equation (\ref{EFE}) gives a modified Friedmann equation
\begin{equation}\label{mfe11}
H^2=\frac{\beta}{8} \left(H^4+\frac{8 Q^2}{3 L^4 a^2}H^2+\frac{36 L^2 M  +8 Q^4}{9 L^8 a^4}\right)+\frac{8\pi G_4}{3}\rho+\frac{\Lambda_4}{3}.
\end{equation}
The $ij$-component of (\ref{EFE}) gives
\begin{equation}\label{mfe12}
\dot{H}+H^2=\frac{\beta}{8}\left(H^4+2\dot{H}H^2+\frac{8Q^2}{3a^2}\dot{H}-\frac{36 L^2 M+8 Q^4}{9 L^8 a^2}
\right)-\frac{4\pi G_4}{3}(\rho+3p)+\frac{\Lambda _4}{3},
\end{equation}
where $\beta=G_4/G_5$. It is easy to check that equations (\ref{mfe11}) and (\ref{mfe12}) are consistent. In the $Q$ $\rightarrow$ $0$ limit, the scalar field and the gauge field vanish and the potential becomes a cosmological constant in the five dimensional AdS. This means that the hairy black hole geometry reduces to the AdS-Schwarzschild black hole geometry. The modified Friedmann equation (\ref{mfe11}) should also reduce to the one derived from the five-dimensional AdS-Schwarzschild black hole. The relevant equation has been derived in \cite{Apostolopoulos:2008ru} (note that there is a typo in the original paper) and is found to be the same with the (\ref{mfe11}) in the $Q$ $\rightarrow$ $0$ limit. Terms proportional to $\beta$ on the right hand side of each equation are from the conformal field theory dual to the gravity. If there are no such terms, the equation falls into the standard expression, the Friedman equation in the $\Lambda$CDM model. 

One last thing worthwhile to mention here is the temperature of the universe. As the black hole in the bulk has the finite Hawking temperature $T_H$, this contributes to the boundary temperature. As the radial coordinates was scaled when we choose the boundary, the temperature of the universe is also scale by the scale factor $a(V)$. So the temperature of the universe has additional $T_H/a(V)$ term. The overall temperature of the boundary attributes to the bulk black hole and the real particles on the boundary.

%%%%%%%%%%%
\section{Fitting models to the observational data}\label{constraint} 

In this section, using the modified Friedmann equation~(\ref{mfe11}) and numerical techniques developed in~\cite{Basilakos:2016nyg}, we test our model in (\ref{eq:Action}) against the observational data and present our results of MCMC analysis.  
\subsection{Models}
The evolution of the universe for our model is governed by the modified Friedmann equation (\ref{mfe11}), which can be rewritten as %(\ref{adssch1}) 
{\small \begin{equation}
\frac{H^4}{H_0^4}-\left[\frac{8}{\beta H_0^2}-\frac{8 Q^2}{3 L^4 H_0^2}(1+z)^2\right]\frac{H^2}{H_0^2}+ \left(\frac{36L^2 M+8Q^4}{9L^8 H_0^4}\right)(1+z)^4+\frac{8}{\beta H_0^2}\left[\Omega_r(1+z)^4+\Omega_m (1+z)^3+\Omega_\Lambda\right]=0\,,
\end{equation}}
where $a(\tau)=a(\tau_0)/(1+z)$ is used.
The solution for above equation is
{\small \begin{equation}\label{eq:H2ofz}
\frac{H^2}{H_0^2}=\frac{1}{2\Omega_\beta}\left(1-\Omega_Q(1+z)^{2} \pm \sqrt{\left(1-\Omega_Q(1+z)^{2}\right)^2-4\Omega_\beta\left[\tilde{\Omega}_r (1+z)^{4}+\Omega_m (1+z)^{3}+\Omega_\Lambda\right]}\,\right)\,,
\end{equation}}
where $\tilde{\Omega}_r\equiv \Omega_r+\Omega_M$ and  the density parameters are defined as 
\begin{equation}
\Omega_\beta\equiv \frac{\beta H_0^2}{8}\,, \,\,\, \Omega_Q\equiv\frac{8 Q^2}{3 L^4 H_0^2}\Omega_\beta\,, \,\,\, \Omega_M\equiv \frac{36L^2M+8 Q^4}{9L^8 H_0^4}\Omega_\beta\,, \,\,\, \Omega_{m,r}\equiv\frac{8\pi G_4 \rho_{m,r}^0}{3 H_0^2}\,,\,\,\, \Omega_\Lambda\equiv\frac{\Lambda_4}{3 H_0^2}\,,
\end{equation}
and $H_0$ is the current value of the Hubble parameter, $H_0=100 h\, \text{km}\, \text{s}^{-1}\, \text{Mpc}^{-1}$. By taking (\ref{eq:H2ofz}) at $z=0$, \emph{i.e.,} $H_0\equiv H(z)|_{z=0}$, we obtain a relation between different energy components as follows
\begin{equation}
1=\frac{1}{2\Omega_\beta}\left[ (1- \Omega_Q)  \pm \sqrt{\left(1 - \Omega_Q\right)^2 - 4\Omega_\beta \left(\Omega_M+\Omega_r+\Omega_m+\Omega_\Lambda\right)}\,\right]\,.
\end{equation}
We obtain from the last equation that
\begin{equation}
\Omega_\Lambda=1-\Omega_m-\tilde{\Omega}_r-\Omega_Q - \Omega_\beta \,,
\end{equation}
where we introduced $\tilde{\Omega}_r$ due to the fact that they have the same evolution hence the number of free parameter reduces by one.

\subsection{Data} 

In our numerical analysis, we use two different observational data sets from the low-redshift measurements including the Supernovae Type Ia (SnIa) and the direct measurements of the Hubble expansion rate. In particular, we use the Pantheon compilation of SnIa data~\cite{Scolnic:2017caz} and %For more statistical analysis of heterogeneity in SnIa, one may refer to \cite{Alam:2017svz}. %We obtain $\chi^2_{SN}$ using JLA sample by marginalizing over $\mathcal{M}$ (nuisance parameter representing some combinations of the absolute magnitude of a fiducial SnIa) following \cite{Conley:2011ku, Basilakos:2016nyg}. 
%We also use the CMB shift parameters, flat prior only, obtained in \cite{Wang:2015tua, Elgaroy:2007bv} based on the \textit{Planck 2015} release~\cite{Ade:2015xua}. 
%Finally, we use  the local measurement of the Hubble constant derived from a re-analysis of Cepheid data~\cite{Riess:2011yx, Efstathiou:2013via}. %where $\chi^2_{H_0}$ is given by $\chi^2_{H_0}=\left[(h-0.706)/0.033\right]^2$. 
the cosmic chronometric data on $H(z)$~\cite{Moresco:2016mzx}. There are two ways of deriving $H(z)$; by the clustering of galaxies or quasars and by the differential age method. The first method provides direct measurements of the $H(z)$ by measuring the BAO peak in the radial direction from the clustering of galaxies or quasars~\cite{Gaztanaga:2008xz} while the second method obtains the $H(z)$ via the redshift drift of distant objects over significant time periods, which is possible as in GR the H(z) can be expressed in terms of change in the redshift, \emph{i.e}., $H(z)=-1/(1+z)dz/dt$~\cite{Jimenez:2001gg}. As a result, these methods provide 36 data points of the $H(z)$ between $0.07\leq z \leq 2.36$. %~\cite{Moresco:2016mzx}. 
%This joint analysis includes the $H_0$ measurement as motivated in~\cite{Xu:2016grp} in order to study different dark energy models. 
%We then perform a MCMC analysis, following~\cite{Basilakos:2016nyg},  
We then compute the total likelihood function $\mathcal{L}_\text{tot}$, which can be written as the product of likelihood functions of each data set,  
%\begin{equation}
$\mathcal{L}_\text{tot}=\mathcal{L}_\text{SnIa}%\times\mathcal{L}_\text{CMB}
\times\mathcal{L}_{H_0}\,.
$ %\end{equation}
The likelihood function can be converted into the sum of the total $\chi^2$,
%\begin{equation}\label{eq:chi2total}
$ \chi^2_\text{tot}=\chi^2_\text{SnIa}%+\chi^2_\text{CMB}
+\chi^2_{H_0}\,,
$ %\end{equation}
where  $\chi^2_\text{tot}=-2\log \mathcal{L}_\text{tot}$ is used. In the following, let us explain the data sets used in the likelihood analysis.

\subsection{MCMC analysis and model comparison}
\begin{figure}[h!] %[htbp]%
\centering  %\subfigure[hFRW]
%{
\includegraphics[width=1.2\textwidth]{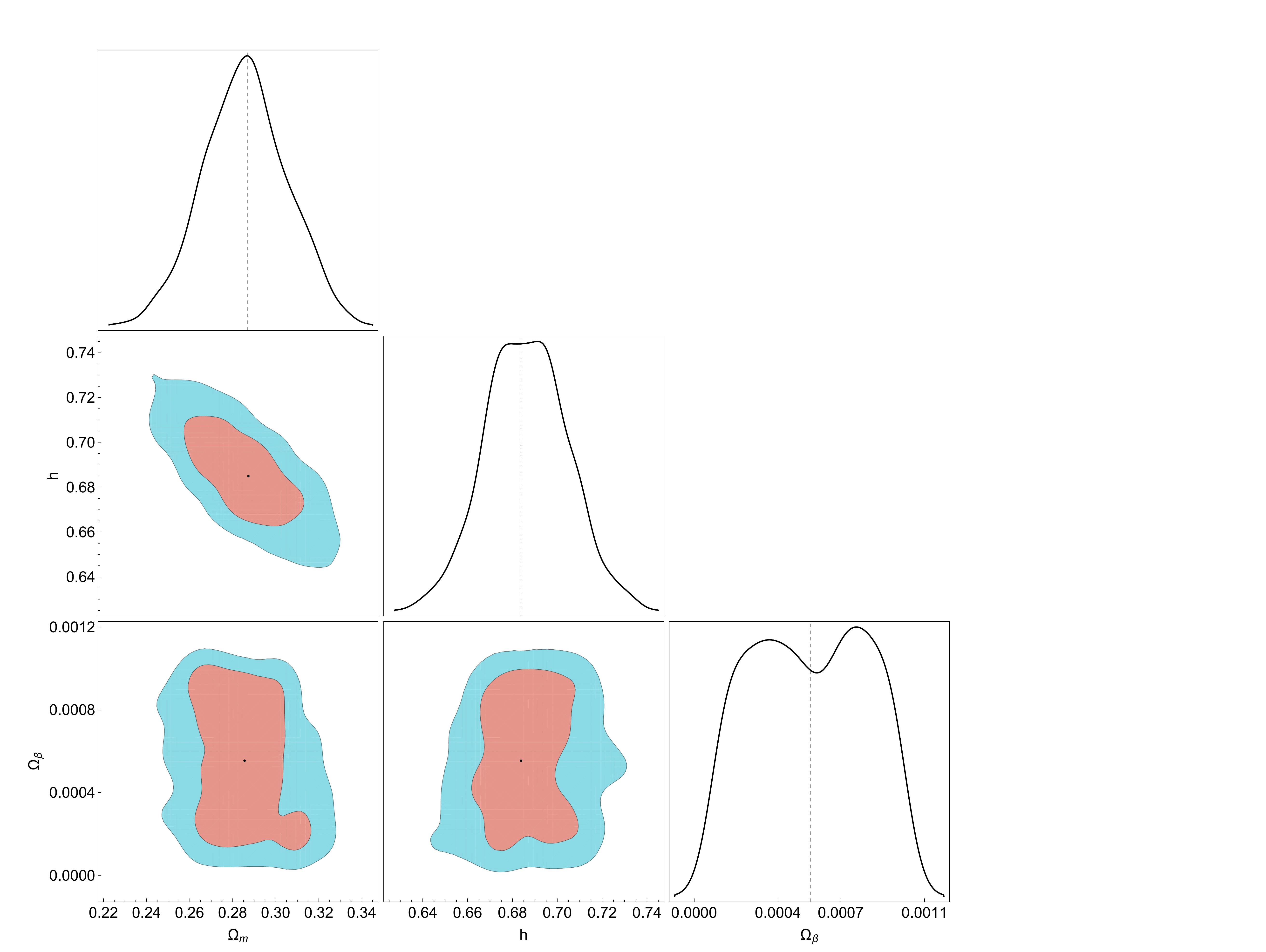} 
\caption{The $68.3\%$ and $95.4\%$ confidence contours between parameters for the Hairless BH case and their 1D marginalized likelihood. The vertical dashed lines and black dots indicate the mean MCMC values at ($\Omega_m$, $h$, $\Omega_\beta$) = ($0.2855\,, 0.6837\,, 5.5431\times10^{-4}$). }\label{fig:hairless1}
\end{figure}

\begin{figure}[h!] %[htbp]%
\centering  
\includegraphics[width=1.2\textwidth]{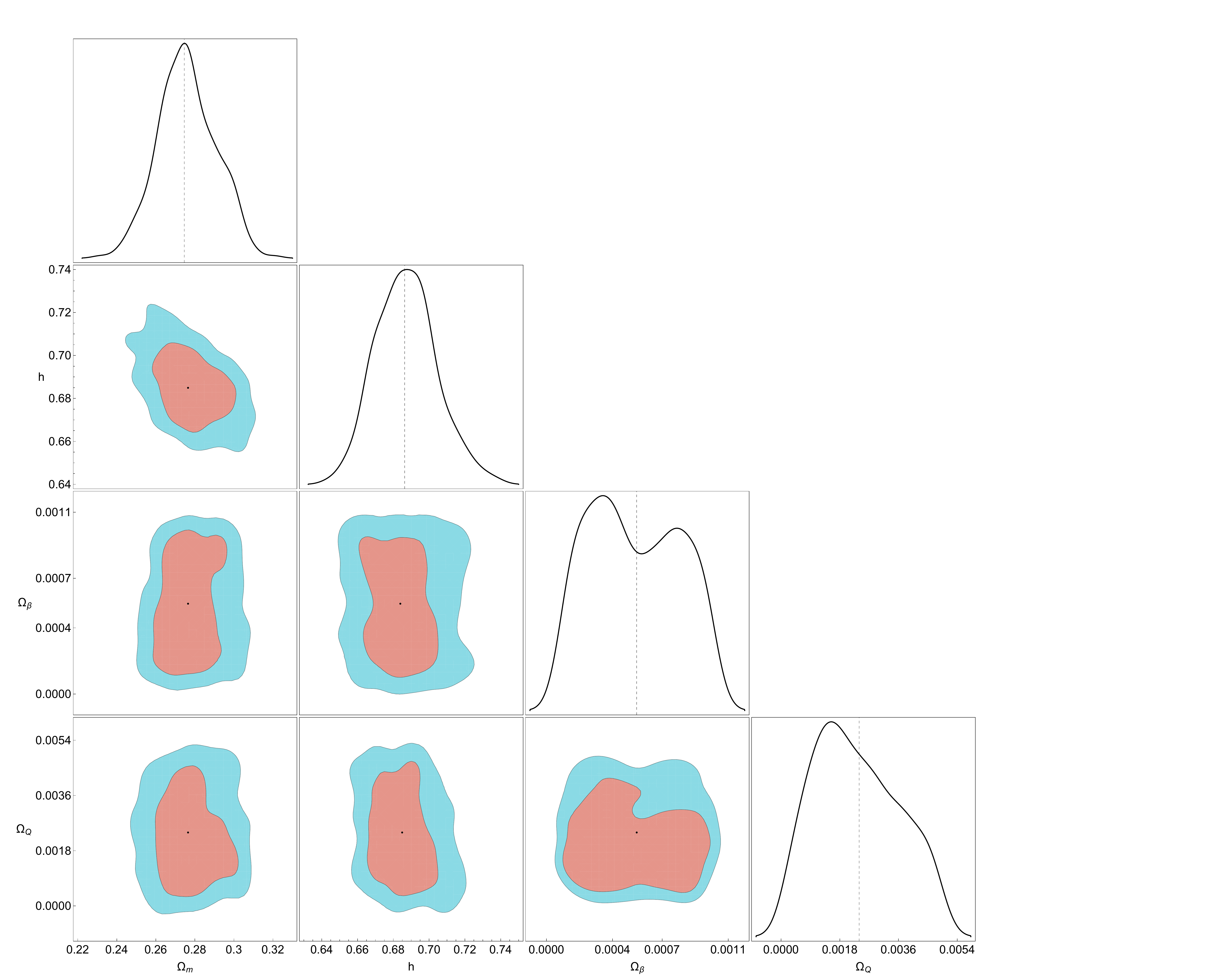} 
\caption{The $68.3\%$ and $95.4\%$ confidence contours between parameters for the Hairy BH case and their 1D marginalized likelihood. The vertical dashed lines and black dots indicate the mean MCMC values at ($\Omega_m$, $h$, $\Omega_\beta$, $\Omega_Q$) = ($0.2765\,, 0.6849\,, 5.4624\times 10^{-4}\,, 2.3937\times 10^{-3}$). }\label{fig:hairy1}
\end{figure}

By using on the $\chi^2$ functions for each data set, we perform a MCMC sampling analysis for the cosmological parameters including $\Omega_m$, $\Omega_b h^2$, $h$, $\Omega_\beta$, and $\Omega_Q$.  We plot one-dimensional probability distribution and two-dimensional observational contours in Figs.~\ref{fig:hairless1} and~\ref{fig:hairy1} for both hairless and hairy BH cases, respectively. The main results regarding the best-fit values are listed in Table~\ref{tab:table1} in comparison to that of the $\Lambda$CDM model. 

\begin{table}[h!]
\centering  %\label{tab1} %$\Omega_b h^2$   & $0.0986\pm0.2152$      & ---   \\ 
\caption{The best-fit values of cosmological parameters and their uncertainties with 68.3\% C.L. \label{tab:table1}} % Results of 
\begin{tabular}{c c c c c c c c c c}
 \hline\hline 
Parameters          &\vline&  $\Lambda$CDM         &\vline& {Hairless BH model} &\vline& {Hairy BH model}    \\ [0.5ex] \hline
$\Omega_m$          &\vline&  $0.2848\pm 0.0187$     &\vline& $0.2872\pm 0.0191$   &\vline& $0.2852\pm 0.0141$  \\  
%$\Omega_b h^2$      &\vline&  $0.0222\pm 0.0001$   &\vline& $0.2222\pm 0.0001$  &\vline& $0.0221\pm 0.0001$  \\
$h$                 &\vline&  $0.6853\pm 0.0169$      &\vline& $0.6849\pm 0.0181$   &\vline& $0.6855\pm 0.0143$  \\    
$\Omega_\beta$      &\vline&      ---              &\vline& $(5.5431\pm 2.6235)\times 10^{-4}$   &\vline& $(5.1275 \pm 2.6600) \times 10^{-4}$  \\ 
$\Omega_Q$          &\vline&      ---              &\vline&     ---             &\vline& $(2.3937\pm 1.2845)\times 10^{-3}$  \\  \hline
%$\alpha$            &\vline&  $0.1403 \pm 0.0068$  &\vline& $0.1367\pm0.0073$   &\vline& $0.1412\pm 0.0069$  \\ 
%$\beta$             &\vline&  $3.1081\pm 0.0892$   &\vline& $3.1256\pm0.0922$   &\vline& $3.0895\pm 0.0886$  \\  \hline
$\chi^2_\text{min}$ &\vline&  $1056.56$            &\vline& $1056.62$           &\vline& $1056.81$           \\  
$\Delta$AIC         &\vline&       0               &\vline& $2.06$             &\vline& $ 4.25 $             \\  
$\Delta$BIC         &\vline&       0               &\vline& $7.01$            &\vline& $ 14.15 $           \\  
\hline\hline
\end{tabular}
\end{table}
The $\Lambda$CDM model, which is currently regarded as the best cosmological model explaining the observational data among all existing ones, can be recovered in our study when $\Omega_\beta=0$. Thus, our models on the four-dimensional conformal boundary of the $AdS_5$ BH can be treated as a simple extension to the $\Lambda$CDM model hence the direct comparison between the models can be done.  

\begin{figure}[h!] %[htbp]%
\centering  
\includegraphics[width=0.6\textwidth]{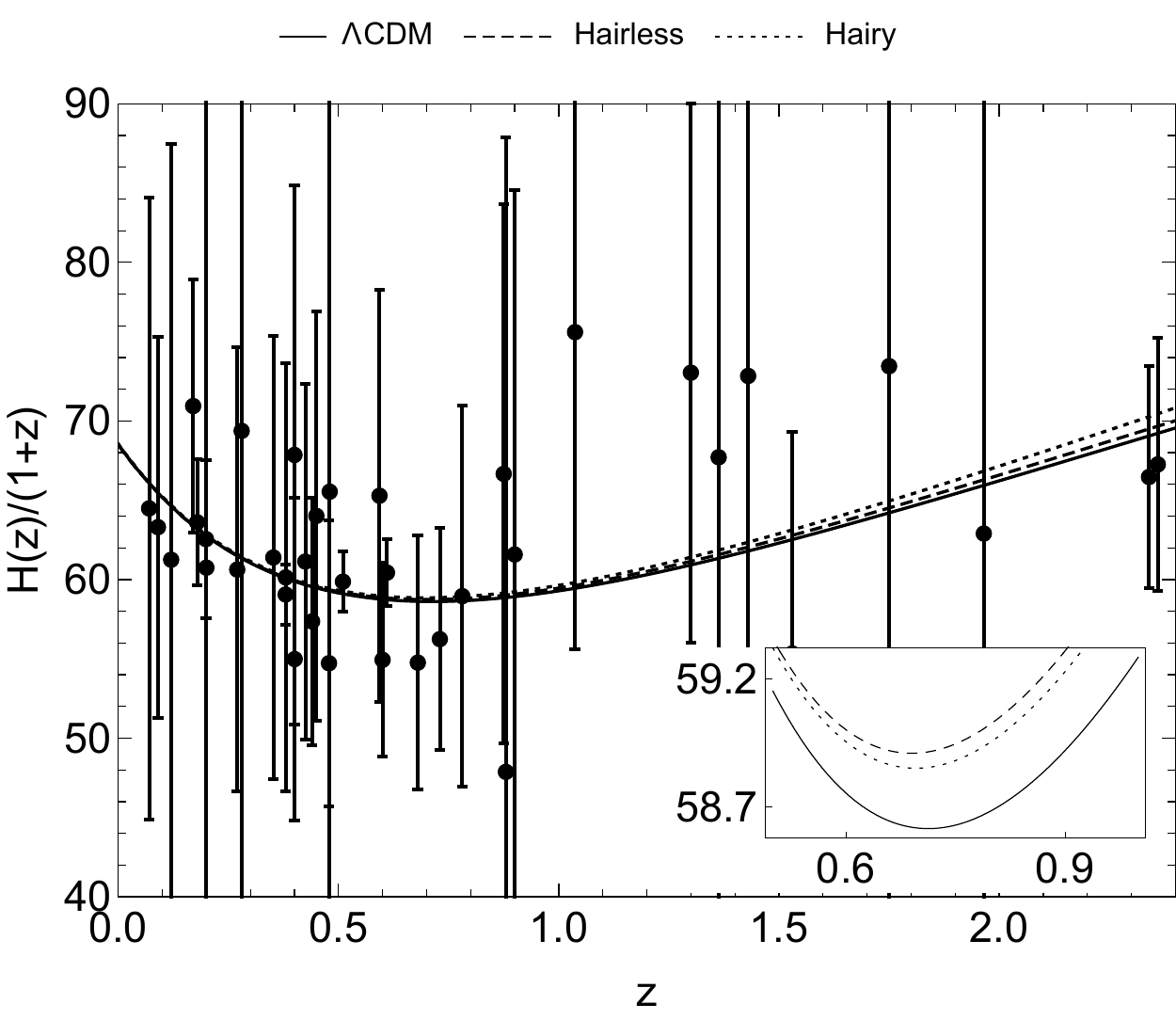} 
\caption{The redshift evolution of~(\ref{eq:H2ofz}) in light of the observational data~\cite{Moresco:2016mzx}, where the numerical inputs for each model are from Table~\ref{tab:table1}. The $\Lambda$CDM model with $H(z) = H_0\sqrt{1- \Omega_m+\Omega_m(1+z)^3}$ is reflected in the solid black line.}\label{fig:compared}
\end{figure}

In regard to comparing the statistical significance our model with the $\Lambda$CDM model, $\chi_\text{min}^2$ cannot make a fair comparison because of the fact that a model with more parameters has more tendency to have a lower value of $\chi_\text{min}^2$ in general. Compared to the $\Lambda$CDM model, our models have one additional parameter ($\Omega_\beta$) for the hairless BH case and two ($\Omega_\beta$ and $\Omega_Q$) more for the hairy BH case. Thus, in order to make a fair comparison, we use well known Akaike information criterion (AIC)~\cite{Akaike1974} and Bayesian information criterion (BIC)~\cite{Schwarz:1978tpv} in our study. 

The AIC and BIC estimators are defined as $ \text{AIC} \equiv -2 \ln \mathcal{L}_\text{max} + 2k$ and $\text{BIC} \equiv -2\ln \mathcal{L}_\text{max} + k \ln N$, where $\mathcal{L}_\text{max}$, $k$, and $N$ indicate the maximum likelihood, the number of free parameters, and the number of data points we use in our model-to-data fitting, respectively. Assuming Gaussian errors, one can use $\chi^2_\text{min}=-2\ln \mathcal{L}_\text{max}$.
The usual interpretation of the AIC and BIC estimator is that a model with a smaller AIC value means a better model in terms of data fitting, while a smaller BIC value indicates that such a model is economically favorable if further data points are implemented. The $\Lambda$CDM is used as a reference model in our study. Thus, we need to use the pair difference between our model and $\Lambda$CDM model; $\Delta \text{AIC}=\text{AIC}_\text{our model}-\Delta \text{AIC}_{{\Lambda\text{CDM}}}$ and  $\Delta \text{BIC}=\text{BIC}_\text{our model}-\Delta \text{BIC}_{{\Lambda\text{CDM}}}$. This can be translated as $\Delta \text{AIC}=\Delta \chi^2_{\text{min}}-2\Delta k$ and $\Delta \text{BIC}=\Delta \chi^2_\text{min}-\Delta k \text{ln}N$, respectively. 

%of the additional free parameters that we have for our model. A 

%When it comes to the model-to-data fitting, this relative difference can be interpreted with the Jeffreys' scale as follows: 
The $\Delta$AIC and $\Delta$BIC can be interpreted similarly to the $\chi_\text{min}^2$, \emph{i.e.,} a relative value signifies a better fit to data. In other words, this relative difference can be interpreted with the Jeffreys' scale as follows: $0<\Delta \text{AIC}\leq 2$ indicates the consistency between two models, $4<\Delta \text{AIC}<7$ suggests a positive evidence against the model with higher value of $\text{AIC}_\text{model}$, and $\Delta \text{AIC}>10$ can be interpreted as an indication of essentially no support with respect to the reference model. For the BIC, the relative difference $\Delta \text{BIC}=\text{BIC}_\text{model}-\text{BIC}_{\Lambda\text{CDM}}$ provides the following situations: $\Delta\text{BIC}\leq 2$ indicates that the model of interest is consistent with the reference model, $2\leq \Delta\text{BIC}\leq 6$ implies the positive evidence against the model, and $\Delta \text{BIC}\geq 10$ suggests that such evidence becomes strong.%~\footnote{It is worth noting that the Jeffreys' scale in general has been shown to lead to misleading conclusion hence one needs to be careful with using this scale.}

%\section{A model comparison using information criterion}

Let us highlight key results of our study in the following. In Figs.~\ref{fig:hairless1} and \ref{fig:hairy1}, we show the $68.3\%$ and $95.4\%$ confidence contours for the hairy and the hairless BH models, respectively, along with the 1D marginalized likelihood for various parameter combinations. %In these plots, we highlight the $\Lambda$CDM model with either a red point or a red dashed line, which correspond to those values listed in the first column of the Table~\ref{tab:table1}. 
The figures, as well as the table, seem to show that our models explain the observational data as good as the $\Lambda$CDM model does. Moreover, as is seen in Table~\ref{tab:table1}, we find that the relative difference $2<\Delta\text{AIC}<7$ and which suggests a positive evidence against our model (both hairy and hairless cases). However, if we take the smallness of AIC and BIC into an account, $\Lambda$CDM model is still favored over our model. The $\Delta\text{BIC}$ values presented in Table~\ref{tab:table1} indicate that, if more data is used, $\Delta$AIC between the two models might be increasing in some extent. Thus, more data can tell us how well these models relatively fit the observational data. %Thus, only the future data can tell us more about how well these models relatively fit the observational data. 

In Fig.~\ref{fig:compared}, and using the best-fit values from Table~\ref{tab:table1}, we plot the low-redshift evolution of the Hubble parameter in our models (\ref{eq:H2ofz}). As is seen in the figure, our models can explain the observational data~\cite{Moresco:2016mzx} as good as the $\Lambda$CDM model does in the redshift $0\leq z \leq 2.36$  interval, and the deviation from the $\Lambda$CDM model noticeable in the redshift increasing direction. We also find the redshift $z_{da}$ of the cosmological deceleration-acceleration transition at $z_{da}^{\Lambda\text{CDM}}\simeq0.7125$, $z_{da}^{\text{Hairless}}\simeq0.7091$, and $z_{da}^\text{Hairy}\simeq 0.6954$ for each cosmological models we discuss in this study. Here, the $z_{da}$ indicates the time that our universe transitioned from non-relativistic (baryon and cold dark) matter dominated phase to the current dark energy dominated phase; hence, $\ddot{a}=0$ at $z_{da}$. The result in Fig.~\ref{fig:compared} indicates that our universe entered the phase of cosmic acceleration slightly later in the holographic models than the $\Lambda$CDM model. Thus, in order to explain the current observational data as good as the $\Lambda$CDM model does, our two holographic models develop a faster expansion rate, a larger $H(z)$, at each redshift after the time of deceleration-acceleration transition; hence a longer history of our Universe to the CMB time. 

\section{Conclusion}\label{conclusion}

We studied the background cosmological evolution from the four-dimensional boundary of a five-dimensional Anti-de Sitter ($AdS_5$) black holes in this work. The four-dimensional conformal boundary takes the FRW geometry with a scale factor, $a(\tau)$. To see the FRW spacetime on the boundary, we employed so-called EF coordinates and used the scale-invariant property of the bulk geometry. %However, the resulting four-dimensional boundary in this setup is known to be the Einstein static universe. Thus, to achieve an expanding universe on the conformal boundary, we expanded the spacetime metric near boundary and performed the coordinate transformation to FG coordinates such that the new cosmological boundary has a desired FRW form of the spacetime metric. 

Modified Friedmann equations, which is our main result of this work, are derived on the FRW boundary of an $AdS_5$ BH through the AdS/CFT correspondence and its background evolution has further investigated. Since the late-time accelerating universe is our main concern here, we treated the extra contributions coming from the bulk side as dark energy and performed MCMC analysis using observational data.  Compared to  the $\Lambda$CDM, our models contain additional free parameters that are associated to the charge $Q$ and mass $M$ of the BH. Thus, to make a fair comparison, we have used AIC and BIC in our analysis.  Albeit the forms of equations we derived look far different from that in the standard model of cosmology, the cosmological evolution of the universe for our model found to be similar to that of the $\Lambda$CDM model. The connection with braneworld models can be found in Ref.~\cite{yunlong:2021}

The key results of our numerical work are presented in Table~\ref{tab:table1} and Figs.~\ref{fig:hairless1}--\ref{fig:compared}. The figures, as well as the table, have shown that our models explain the observational data as good as the $\Lambda$CDM model does for the current data. However, if we take the smallness of $\Delta$AIC and $\Delta$BIC into an account, $\Lambda$CDM model is still favored over our model. Moreover, the $\Delta\text{BIC}$ values presented in Table~\ref{tab:table1} indicate that, if more data is used, $\Delta$AIC between the two models might be increasing in some extent. Thus, more data can tell us how well these models relatively fit the observational data. 

%Modified Friedmann equations are obtained on the four-dimensional conformal boundary of a five dimensional AdS black holes through $AdS/CFT$ correspondence. 

\section{Acknowledgement}

We thank Yoobin Jeong for his contribution at the initiation of the project and Yun-Long Zhang for his helpful discussions and valuable comments on an earlier version of the manuscript. SK was supported by Higher Education Improvement Project (HEIP), 6221-KH. BHL was supported by Basic Science Research Program through the National Research Foundation of Korea(NRF) 2020R1A6A1A03047877 and also by 2020R1F1A1075472. GT was supported by Ministry of Science and Technology (MoST) grant No. 109-2112-M-002-019.

\end{document}